\documentclass[12pt]{article}
\usepackage{amsmath}
\usepackage{amssymb}
\def\t           {\tilde}
\def\thetabar    {\bar{\theta}}
\def\g           {\mathfrak{g}}

\def\N           {\mathbb N}
\def\Z           {\mathbb Z}

\def\C           {\mathbb C}

\def\fsl         {sl(2|1;\mathbb C)}

\def\hslc        {\hat{sl}(2|1;{\mathbb C})}
\def\hslck       {\hat{sl}(2|1;{\mathbb C})_k}

\def\hf          {\tfrac{1}{2}}
\def\thf         {\tfrac{3}{2}}

\begin{document}

\numberwithin{equation}{section}

\begin{titlepage}
\begin{flushright}
DCTP-01/45\\
April 2001\\
\end{flushright}
\vspace{1cm}
\begin{center}
{\Large\bf Fusion Rules for $\hslc$ at Fractional Level $k=-1/2$}\\
\vspace{1cm}
{\large Gavin Johnstone}\\
\vspace{0.5cm}
{\it Department of Mathematical Sciences, University of Durham, \\Durham DH1 3LE, England}\\
\vspace{0.5cm}
e-mail: G.B.Johnstone@durham.ac.uk
\end{center}
\vspace{2cm}

\begin{abstract}

We calculate fusion rules for the admissible representations of the
superalgebra $\hslck$ at fractional level $k=-1/2$ in the Ramond sector.  
By representing 3-point correlation functions involving a singular
vector as the action of differential operators on the $sl(2|1;\C)$ invariant
3-point function, we obtain conditions on permitted quantum
numbers involved.  We find that in this case the primary fields close
under fusion.

\end{abstract}

\end{titlepage}

\section{Introduction}

The study of conformal field theory based on affine algebras is one
that is well-established.  Unitary theories at integer level do not,
however, provide the whole story, and it is interesting to also
consider theories built on affine algebras at fractional level: these
may provide access to many other models via hamiltonian reduction, or
indeed (algebraically) define non-unitary models in 
their own right.  In this regard, it is admissible representations
which stand out, since their characters form a closed set under
modular transformations.  The case of $\hat{sl}(2)$ is much studied:
in particular, fusion rules for fractional level were investigated in
\cite{MW, AY, BF, FGP}, with more abstract analysis carried out in
\cite{FM} and \cite{DLM}. This question has also been
addressed in the case of $\hat{sl}(3)$ in \cite{FGP1, FGP2} and 
in \cite{GPW}, where an overview of the $\hat{sl}(2)$ situation is also given.  
Conformal blocks for fractional level $\hat{sl}(2)$ have also been studied
extensively, for example in \cite{FGP, PRY, And}.  As for
superalgebras, the case of $\widehat{osp}(1|2)$ has been considered in
\cite{ER}.  It is clear then that the subject of fractional level is
one of continued interest, the present work concerned with extending
the techniques of \cite{ER} to discuss the superalgebra $\hslc$.  This
introduces an additional degree of complexity, arising from the
fact that $\hslc$ is the simplest superalgebra where zero length
roots appear, in contrast to $\widehat{osp}(1|2)$.  It is worthwhile
studying this from an abstract point of view, but also given the
intimate link between non-unitary $\hslc$ and the $N=2$ non-critical
superstring \cite{Sem}.  Fractional level $\hslc$ has also arisen in
the study of Gaussian disordered systems \cite{MS}. 

Common to most of the works mentioned above is the understanding that
for fractional level representations, one must work with fields which
are not only a function of the usual coordinate $z$, but also of an
isotopic coordinate $x$, representing an internal $sl(2)$ symmetry.
This technique was first applied to the unitary $\hat{sl}(2)$ case in
\cite{FZ} and is also developed in \cite{BS}.  For fractional level,
this overcomes the problem of needing to consider general
infinite-dimensional representations, which are neither highest nor
lowest weight.  In \cite{ER}, the authors extended this approach by
including not only the coordinate $x$ but also dependence on a
Grassmann coordinate $\theta$ to represent the supersymmetry present
in $\widehat{osp}(1|2)$: for the case of $\hslc$ we must additionally
augment this by another Grassmann coordinate $\thetabar$, now with two
supersymmetries present.  Our basic approach will be to determine
3-point correlation functions involving fields with this
dependence and then determine this correlator involving a singular
vector.  We may then rewrite this as an expression involving
differential operators acting on the 3-point function, which must
equal zero since it involves a singular vector.  This will provide
relations between the quantum numbers of the three fields present,
determining which fusion rules are permitted.  

We begin by recalling some essential features of $\hslc$.  We then
discuss the relation between highest weight states and primary fields,
before going on to determine a realisation of $\hslc$ in terms of
differential operators.  This will allow us to determine the $\hslc$
invariant 3-point function, an issue discussed in the context of
$N=2$ superconformal field theory; in examining this issue we will
relate the work of \cite{Kir, MSS, West, Bl} on this subject.  A
brief summary of singular vectors required (as determined in
\cite{BT97}) will then be given, before putting these pieces together
in the calculation of fusion rules.

\section{Introduction to $\hslc$}

We begin with a brief overview of some of the important features of
$\hslc$.  For more details, the references \cite{BT97}, \cite{BHT98}
may be consulted. 

The affine superalgebra $\hslck$ is made up of the even generators
$\{J^{\pm}_n,J^3_n,U_n\}$ and odd generators
$\{j^{\pm}_n,j^{'\pm}_n\}$, supplemented by the usual affine
generators $\t{k}$, the central generator, and $d$, the derivative
operator.  This is identified with the generator $-L_0$ of the
Virasoro algebra associated to $\hslck$ via the Sugawara construction.
The non-zero (anti)commutation relations of $\hslck$ are 
\begin{xalignat}{2}
\label{cr}
[J_m^+,J_n^-]&= 2J_{m+n}^3+2\t{k}m \delta_{m+n,0},&[U_m,U_n]&=
-\t{k}m \delta_{m+n,0},\notag\\
[J_m^3,J_n^{\pm}] &= \pm
J_{m+n}^{\pm},&[J_m^3,J_n^3]&=\t{k}m \delta_{m+n,0},\notag\\
[J_m^{\pm},j_n^{'\mp}]&=\pm j_{m+n}^{\pm},&[J_m^{\pm},j_n^{\mp}]&=\mp
j_{m+n}^{'\pm},\notag\\
[2J_m^3,j_n^{'\pm}]&=\pm j_{m+n}^{'\pm},&[2J_m^3,j_n^{\pm}]&=\pm
j_{m+n}^{\pm}\notag,\\
[2U_m,j_n^{'\pm}]&=\pm j_{m+n}^{'\pm},&[2U_m,j_n^{\pm}]&=\mp
j_{m+n}^{\pm},\notag\\
\{j_m^{'+},j_n^{'-} \}&= (U_{m+n}-J_{m+n}^3)-2\t{k}m
\delta_{m+n,0},&\notag\\
\{j_m^+,j_n^- \}&= (U_{m+n}+J_{m+n}^3)+2\t{k}m \delta_{m+n,0}, &\{
j_m^{'\pm},j_n^{\pm} \} &= J_{m+n}^{\pm}.
\end{xalignat}
In addition, we have that
\begin{equation}
[d,X_n]=nX_n
\end{equation}
and the central generator $\t{k}$ commutes with all other generators.

The even generators have mode index $n \in \Z$, whereas the odd
generators have $n \in \Z$ in the Ramond sector and $n \in \Z+\hf$ in
the Neveu-Schwarz sector.  Setting this index to zero recovers the
finite $\fsl$ algebra.  Unless otherwise stated, we work in the
Ramond sector.

As usual, we can define a triangular decomposition of the algebra via
the generators $d$ and $J_0^3$, combined as the principal gradation
$\t{d}=ad(J_0^3)+ad(d)$:

\begin{align}
\t{d}(J_n^{\pm})&=2n \pm 1, \quad \t{d}(J_n^3)=2n, \quad
\t{d}(j^{\pm}_n)=2n \pm \hf,\\
\t{d}(j^{' \pm}_n)&=2n \pm \hf, \quad \t{d}(U_n)=2n, \quad
\t{d}(\t{k})=\t{d}(d)=0.
\end{align}
Denoting the algebra by $\hat{\g}$, we obtain the decomposition 
\begin{equation}
\hat{\g}=\hat{\g}_- \oplus \hat{\g}_0 \oplus \hat{\g}_+,
\end{equation}
$\hat{\g}_-$ consisting of those elements having $\t{d}<0$,
$\hat{\g}_0$ of elements with $\t{d}=0$ and $\hat{\g}_+$ of
elements with $\t{d}>0$.

Highest weight states  of $\hslck$ are characterised by their
conformal weight $h$, isospin $\hf h_-$ and charge $\hf h_+$:
\begin{equation}
|\Lambda \rangle=|h, \hf h_-, \hf h_+ \rangle 
\end{equation}
where
\begin{equation}
L_0|\Lambda \rangle=h|\Lambda \rangle, \quad J^3_0|\Lambda \rangle=\hf
h_-|\Lambda \rangle, \quad U_0|\Lambda \rangle=\hf h_+|\Lambda
\rangle.
\end{equation}
$|\Lambda \rangle$ is annihilated by all the raising operators, {\em i.e.}
elements of $\hat{\g}_+$.  This condition is equivalent to the
following three: 
\begin{equation}
j_0^+ |\Lambda \rangle = j_0^{'+} |\Lambda \rangle = J_1^- |\Lambda
\rangle =0
\end{equation}
and corresponds to a particular choice of simple roots of $\fsl$. 

The relationship between quantum numbers $h$, $h_-$ and $h_+$ is given by
\begin{equation}
h=\frac{1}{4(k+1)}(h_-^2-h_+^2)
\end{equation}
where $k$ is the level of the representation concerned (taken to be
$\hf$ for the purposes of this work).  The Kac-Khazdan determinant
formula dictates that the Verma module built on a highest weight state
with certain specific $h_-$ and $h_+$ values will contain singular
vectors, a full analysis of which was carried out in \cite{BT97} and
\cite{BHT98}.  We shall make extensive use of the results of these
references for the $k=-\hf$ case.

\section{Fields and States}

In conformal field theory, the link between highest weight states and
primary fields is given by the state-field correspondence, namely
\begin{equation}
\lim_{z, \bar{z} \rightarrow 0} \phi(z,
\bar{z})\,|0 \rangle = |\Lambda \rangle,
\end{equation}
where $|0 \rangle$ denotes the vacuum state of the theory.  It
has been noted by several authors \cite{AY, BF, FGP, FM, DLM, PRY,
And}  that for the case of 
admissible $\hat{sl}(2)$ (and $\hat{sl}(3)$ \cite{FGP1, FGP2, GPW})
representations that dependence on a second $sl(2)$ coordinate should
be added: this overcomes the difficulty in describing these
infinite-dimensional representations, which are in general neither
highest nor lowest weight.  The case of $\widehat{osp}(1|2)$ was
studied in \cite{ER} where additionally a Grassmann coordinate
$\theta$ was introduced to represent the supersymmetry present.  For
the case of $\hslck$, with two supersymmetries, it is necessary to
consider not only the coordinate $z$, but also an additional $sl(2)$
coordinate $x$ as well as two Grassmann coordinates $\theta$ and
$\thetabar$.  This has been formalised in \cite{Ras}.  Anti-holomorphic 
counterparts of all these coordinates
should also be included, although here we suppress any dependence on
these coordinates for clarity.  Thus in our case, the state-field
correspondence is given by
\begin{equation}
\label{stfi}
\lim_{z,x,\theta,\thetabar \rightarrow
0} \phi_a (z,x,\theta,\thetabar)\,|0 \rangle =
|\Lambda_a \rangle.
\end{equation}

We can represent the action of the $\hslc$ modes on primary fields via
the action of certain differential operators, using the commutation
relations 
\begin{align}
\label{fcr}
[J_n^{\alpha},\phi_a (z,x,\theta,\thetabar)]&=z^n D^{\alpha}_a
\phi_a (z,x,\theta,\thetabar),\notag\\
[j_n^{\alpha},\phi_a (z,x,\theta,\thetabar)]&=z^n d^{\alpha}_a
\phi_a (z,x,\theta,\thetabar),\notag\\
[j_n^{'\alpha},\phi_a (z,x,\theta,\thetabar)]&=z^n
d^{'\alpha}_a \phi_a (z,x,\theta,\thetabar),\notag\\
[U_n,\phi_a (z,x,\theta,\thetabar)]&=z^n D^U_a \phi_a
(z,x,\theta,\thetabar).
\end{align}
It should be noted that for the case of the fermionic modes $j$,
$j^{'}$ the commutator should be replaced by an anticommutator as
appropriate.  In addition, we have the relation with Virasoro modes
given by
\begin{equation}
[L_n,\phi_a (z,x,\theta,\thetabar)]=\{ z^{n+1} \partial_z +
(n+1)h^az^n \} \phi_a (z,x,\theta,\thetabar).
\end{equation}
This leads us to the definition of the action of $\phi_a$ on the vacuum:
\begin{equation}
\label{fvac}
\phi_a (z,x,\theta,\thetabar)\,|0 \rangle = e^{zL_{-1}+xJ_0^-
+ \theta j_0^- +\thetabar j_0^{'-}} |\Lambda_a \rangle.
\end{equation}

\section{Differential Operators for $sl(2|1)$}

The equations \eqref{fcr} and \eqref{fvac} allow the calculation of
the explicit forms of the differential operators appearing in
\eqref{fcr}.  Considering the $\hslc$ zero modes, we may write
\begin{equation}
e^{-xJ_0^- - \theta j_0^- - \thetabar j_0^{'-}} X_0 e^{xJ_0^- + \theta j_0^- + \thetabar j_0^{'-}} |\Lambda_a \rangle = e^{-xJ_0^- - \theta j_0^- - \thetabar j_0^{'-}} D^X e^{xJ_0^- + \theta j_0^- + \thetabar j_0^{'-}}|\Lambda_a \rangle,
\end{equation}
where $D^X$ is the differential operator corresponding to the
generator $X_0$.  By repeated differentiation of this expression with
respect to $x$ and application of the relations \eqref{cr} we obtain
the following conjugation formulae:
\begin{align}
e^{-xJ_0^- - \theta j_0^- - \thetabar j_0^{'-}} J_0^3 e^{xJ_0^- +
\theta j_0^- + \thetabar j_0^{'-}}|\Lambda\rangle &= 
J_0^3 -xJ_0^- - \hf \theta j_0^- - \hf \thetabar j_0^{'-}|\Lambda\rangle\notag\\
e^{-xJ_0^- - \theta j_0^- - \thetabar j_0^{'-}} J_0^+ e^{xJ_0^- +
\theta j_0^- + \thetabar j_0^{'-}}|\Lambda\rangle &= 2x J_0^3 -x^2 J_0^- - x \theta
j_0^- - x \thetabar j_0^{'-} + \theta \thetabar U_0|\Lambda\rangle\notag\\
e^{-xJ_0^- - \theta j_0^- - \thetabar j_0^{'-}} J_0^- e^{xJ_0^- +
\theta j_0^- + \thetabar j_0^{'-}}|\Lambda\rangle &= J_0^-|\Lambda\rangle\notag\\
e^{-xJ_0^- - \theta j_0^- - \thetabar j_0^{'-}} j_0^+ e^{xJ_0^- +
\theta j_0^- + \thetabar j_0^{'-}}|\Lambda\rangle &= -xj_0^{'-} + x\theta J_0^- -
\theta (J_0^3 + U_0) + \hf \theta \thetabar j_0^{'-}|\Lambda\rangle\notag\\
e^{-xJ_0^- - \theta j_0^- - \thetabar j_0^{'-}} j_0^- e^{xJ_0^- +
\theta j_0^- + \thetabar j_0^{'-}}|\Lambda\rangle &= j_0^- - \thetabar J_0^-|\Lambda\rangle\notag\\
e^{-xJ_0^- - \theta j_0^- - \thetabar j_0^{'-}} j_0^{'+} e^{xJ_0^- +
\theta j_0^- + \thetabar j_0^{'-}}|\Lambda\rangle &= xj_0^- - x\thetabar J_0^- +
\thetabar (J_0^3 - U_0) + \hf \theta \thetabar j_0^-|\Lambda\rangle\notag\\
e^{-xJ_0^- - \theta j_0^- - \thetabar j_0^{'-}} j_0^{'-} e^{xJ_0^- +
\theta j_0^- + \thetabar j_0^{'-}}|\Lambda\rangle &= j_0^{'-} - \theta J_0^-|\Lambda\rangle\notag\\
e^{-xJ_0^- - \theta j_0^- - \thetabar j_0^{'-}} U_0 e^{xJ_0^- + \theta
j_0^- + \thetabar j_0^{'-}}|\Lambda\rangle &= U_0 + \hf \theta j_0^- - \hf \thetabar
j_0^{'-} - \hf \theta \thetabar J_0^-|\Lambda\rangle.
\end{align}
Comparing these results with that of calculating
\begin{multline}
e^{-xJ_0^- - \theta j_0^- - \thetabar j_0^{'-}} (a\partial_x + b
\theta \partial_x +c \thetabar \partial_x + d\partial_{\theta}
+e\theta \partial_{\theta} +f\thetabar \partial_{\theta}\\
+g\partial_{\thetabar} +h\theta \partial_{\thetabar} +l\thetabar
\partial_{\thetabar}+m\theta\thetabar\partial_{\theta}+n\theta\thetabar\partial_{\thetabar}) e^{xJ_0^- + \theta j_0^- + \thetabar j_0^{'-}}|\Lambda\rangle
=\\
 (a J_0^- + b\theta J_0^- +c \thetabar J_0^- + d j_0^- - \hf d
\thetabar J_0^- +e\theta j_0^- - \hf e\theta \thetabar  J_0^-\\
 + f\thetabar j_0^- + g j_0^{'-} - \hf g\theta J_0^- + h\theta
j_0^{'-} + l\thetabar j_0^{'-} +\hf l\theta \thetabar J_0^- +m\theta\thetabar j_0^- +n\theta\thetabar j_0^{'-})|\Lambda\rangle
\end{multline}
we arrive at the following expressions for the differential operator
realisation of $\fsl$:
\begin{align}
\label{diffop}
D^3 &= -x\partial_x - \hf\theta \partial_{\theta} - \hf \thetabar
\partial_{\thetabar} - \hf h_- \notag\\
D^+ &= -x^2 \partial_x - x\theta \partial_{\theta} - x\thetabar
\partial_{\thetabar} + xh_- + \hf \theta \thetabar h_+ \notag\\
D^- &= \partial_x \notag\\
d^+ &= -x \partial_{\thetabar} + \hf x\theta \partial_x - \hf \theta
(h_- + h_+) + \hf \theta \thetabar \partial_{\thetabar} \notag\\
d^- &= \partial_{\theta} - \hf \thetabar \partial_x \notag\\
d^{'+} &= x \partial_{\theta} - \hf x\thetabar \partial_x + \hf \thetabar
(h_- - h_+) + \hf \theta \thetabar \partial_{\theta} \notag\\
d^{'-} &= \partial_{\thetabar} - \hf \theta \partial_x \notag\\
D^U &= \hf \theta \partial_{\theta} - \hf \thetabar
\partial_{\thetabar} + \hf h_+.
\end{align}

\section{$sl(2|1)$ Invariant 3-point Function}

In order to discover $\hslck$ fusion rules, we wish to consider
quantities such as 
\begin{equation}
\label{schem}
\langle \Lambda^*_1 | \phi_2(z,x,\theta,\thetabar) | \omega
\Lambda_3 \rangle,
\end{equation}
where $\omega\Lambda$ is a singular vector.  Then \eqref{schem} will
be equal to zero and we may rewrite this expression as
\begin{equation}
\label{schem1}
(\textrm{differential operators})\langle \Lambda^*_1 |
\phi_2(z,x,\theta,\thetabar) | \Lambda_3 \rangle = 0
\end{equation}
using the relations \eqref{fcr}.  Evaluating \eqref{schem1} we then
obtain conditions on possible quantum numbers of $\phi_2$ and
$\phi^*_1$ given those of $\phi_3$, amounting to a specification
of allowed fusings.  We see here the appearance of the conjugate field 
$\phi^*_1$, since the fusion rule $\phi_i\times\phi_j=\phi_k$ arises from 
the non-vanishing of the 3-point function $\langle\phi^*_k\phi_j\phi_i\rangle$.
This procedure relies on knowledge of $\hslck$
singular vectors, which we have from \cite{BT97}, and knowledge of the
$\fsl$ invariant 3-point function, which we now move on to
discuss.

The differential operator realisation \eqref{diffop} allows us to
determine the $\fsl$ invariant 3-point function, which is in
fact the 3-point function for $N=2$ superconformal field theory,
since $\fsl$ is isomorphic to the set of generators of the $N=2$
super-M\"{o}bius group.  This has been considered by several authors
\cite{Kir, MSS, West, Bl}, the interrelation of whose work will be
clarified here.   We note here that our discussion
of the Ramond sector directly parallels the discussion of the
Neveu-Schwarz sector in these (and other) works on superconformal
field theory: the algebra \eqref{cr} has vanishing central piece for the
Ramond sector zero mode subalgebra, whereas the equivalent subalgebra
in the usual superconformal discussion is in the Neveu-Schwarz
sector.  Indeed, there it is Neveu-Schwarz generators which give rise to
the super-M\"{o}bius group, as discussed in \cite{West}.  Although our
terminology derives from the fact that we are considering the
situation where the fermionic modes have integer labels, when we
discuss Ramond fields and states they are more like Neveu-Schwarz
rather than Ramond, in the sense that the fields do not introduce branch
cuts in the operator product expansion with fermionic currents.  This is 
due to the fact that our fermionic currents are expanded as, for example 
(see \cite{BHT98}), 
$J({\bf e_{\pm \alpha_1}})(z)= \sum_n j_n^{'\pm}z^{-n-1}$ whereas a typical 
fermionic current in superconformal field theory is $G(z)=\sum_n G_n 
z^{-n-3/2}$.  As we
will find that the fusion of two Ramond fields gives rise to another
Ramond field, this interpretation means that our results are not in
conflict with those of (for example) \cite{Watts} for the $N=1$
superconformal case and \cite{MSS} for $N=2$, where it was found
that the fusion of two Ramond fields produces a Neveu-Schwarz field,
whereas the fusion of two Neveu-Schwarz fields gives another
Neveu-Schwarz field.  
 
For a 3-point function to be $\fsl$ invariant, we require
that
\begin{equation}
\langle 0 | [X_0,\phi_1] \phi_2 \phi_3 | 0 \rangle + \langle
0 | \phi_1 [X_0,\phi_2] \phi_3 | 0 \rangle + \langle 0
| \phi_1 \phi_2 [X_0,\phi_3] | 0 \rangle =0
\end{equation}     
for each of the $\fsl$ generators $X_0$.  Using the relations
\eqref{fcr} yields
\begin{equation}
\label{des}
\sum_{i=1}^3 D^X_i \langle 0 | \phi_1 (z_1,x_1,\theta_1,\thetabar_1)
\phi_2 (z_2,x_2,\theta_2,\thetabar_2)
\phi_3(z_3,x_3,\theta_3,\thetabar_3) | 0 \rangle = 0,
\end{equation}
where $D_i^X$ is the differential operator corresponding to $X_0$,
taking its parameters $h_+$ and $h_-$ from the primary field
$\phi_i$.  This assumes that the vacuum $| 0 \rangle$ is
annihilated by elements of $\fsl$.  As is well known, in conformal
field theory solving the resulting differential equations determines
the 3-point function exactly.  In the $N=2$ superconformal
case, the 3-point function can depend on the nine variables $x_i$,
$\theta_i$ and $\thetabar_i$.  Finding a 3-point function which
satisfies the differential equations arising from the generators
$J_0^+$, $j_0^-$ and $j_0^{'-}$ will result in the automatic
solution of the remaining equations, from the commutation relations
\eqref{cr}.  With nine parameters but only three independent equations
available (although we make use of five equations for
simplicity), there will naturally be some ambiguity in the final answer. 
With this in mind, we proceed with our analysis, closely following the
above references and particularly \cite{West}.  

From the equations for $J_0^-$, $j_0^-$ and $j_0^{'-}$, it is clear
that the 3-point function depends on the following variables:
\begin{equation}
s_{ij} = x_i-x_j-\hf \theta_i\thetabar_j -\hf \thetabar_i \theta_j,
\quad \theta_{ij} = \theta_i - \theta_j, \quad \thetabar_{ij} =
\thetabar_i - \thetabar_j, \quad i,j=1,2,3.
\end{equation}
Then from the $U_0$ equation (suppressing the $z$ dependence for ease
of notation)
\begin{equation}
\sum_{i=1}^3 (\hf \theta \partial_{\theta} - \hf \thetabar
\partial_{\thetabar} + \hf h_+) \langle 0 | \phi_1
(x_1,\theta_1,\thetabar_1) \phi_2  (x_2,\theta_2,\thetabar_2) \phi_3
(x_3,\theta_3,\thetabar_3)| 0 \rangle = 0
\end{equation}
we find that there are several distinct cases to be examined, that is,
for which
\begin{equation}
H_+ = \sum_{i=1}^3 h_{+,i} = 0, \pm 1, \pm 2. 
\end{equation}
Consideration of the $J_0^+$ condition eliminates the cases where
$H_+ = \pm 2$, so there are in fact three distinct
3-point functions to be obtained.  It remains to solve the $J_0^+$
equation subject to the conditions $H_+ = 0, \pm 1$.
For the case $H_+ = 0$, we find (again suppressing the $z$ dependence)
\begin{multline}
\label{3pte}
\langle 0 | \phi_1 \phi_2 \phi_3 | 0 \rangle = C_{123}
s_{12}^{a_3} s_{23}^{a_1} s_{13}^{a_2} \bigg[ 1 +
\frac{h_{+,1} \theta_{12}\thetabar_{12}}{2s_{12}}\\
 + \frac{(h_{+,1}+h_{+,2})
\theta_{23}\thetabar_{23}}{2s_{23}} + \frac{h_{+,1}(h_{+,1}+h_{+,2})
\theta_{12}\thetabar_{12}\theta_{23}\thetabar_{23}}{4s_{12}s_{23}}\\
 + \alpha \frac{\theta_{12}\thetabar_{12}s_{23}}{s_{12}s_{13}} - \alpha
\frac{(\theta_{12}\thetabar_{23} + \theta_{23}\thetabar_{12})}{s_{13}}
+ \alpha \frac{\theta_{23}\thetabar_{23}s_{12}}{s_{23}s_{13}}\\
 + \alpha
\frac{h_{+,1}\theta_{12}\thetabar_{12}\theta_{23}\thetabar_{23}}{2s_{23}s_{13}}+
\alpha
\frac{(h_{+,1}+h_{+,2})\theta_{12}\thetabar_{12}\theta_{23}\thetabar_{23}}{2s_{12}s_{13}}\bigg],  
\end{multline}
where $\alpha$ is an undetermined parameter and
$a_1=\hf(h_{-,2}+h_{-,3}-h_{-,1})$, {\em etc.}  This is essentially the
answer of \cite{MSS} and \cite {Bl}, though now with no restrictions
on $\alpha$.  The system of equations given by
\eqref{des} is invariant under permutations of 1,2 and 3: this is not
reflected in the solution \eqref{3pte} and indeed any permutation of
these labels will also give a solution.  If we interchange 1 and 3 and add
the resulting answer to the one above, we obtain a solution
\begin{multline}
\label{3pte1}
\langle 0 | \phi_1 \phi_2 \phi_3 | 0 \rangle =
2 C_{123} s_{12}^{a_3} s_{23}^{a_1} s_{13}^{a_2} \bigg[ 1 +
\frac{h_{+,1} \theta_{12}\thetabar_{12}}{2s_{12}}
 + \frac{(h_{+,1}+h_{+,2})
\theta_{23}\thetabar_{23}}{2s_{23}}\\
 + \frac{h_{+,1}(h_{+,1}+h_{+,2})
\theta_{12}\thetabar_{12}\theta_{23}\thetabar_{23}}{4s_{12}s_{23}}\bigg],  
\end{multline}
where we have used the fact that $H_+=0$ and taken $C_{123}=C_{321}$.
Strictly speaking, this is  not the exact answer obtained by this
procedure, since $s_{ij}=-s_{ji}$ means that the permuted answer differs from
\eqref{3pte} by a factor $(-1)^{-a_1-a_3-a_2}$.  However, when the
full dependence on anti-holomorphic variables is included, the overall
multiplicative factor in \eqref{3pte} is modified to
$C_{123}|s_{12}|^{-2a_3}|s_{23}|^{-2a_1}|s_{13}|^{-2a_2}$ (with
$h_{-,i}=\bar{h}_{-,i}$) and this discrepancy disappears.  The expression
\eqref{3pte1} is in fact a particular case of the solution obtained by
Kiritsis \cite{Kir}.  Understanding \eqref{3pte} as written for the
labelling \{123\} we find that the solution obtained by adding
\eqref{3pte} written with \{213\} to that with \{312\} is also a particular
Kiritsis solution, as is the expression resulting from the addition of
\{132\} to \{231\}.  When all six versions of \eqref{3pte} are added
together, the solution obtained is precisely that given by Howe and West
\cite{West}, which is again a specific instance of the solution
described by Kiritsis, distinguished by the fact that it is a
permutation invariant solution of the equations \eqref{des}.  Before
going on to clarify this situation, we consider the other 3-point
functions for the cases $H_+=\pm 1$.

When $H_+=-1$, we find that
\begin{multline}
\label{3pto1}
\langle 0 | \phi_1 \phi_2 \phi_3 | 0 \rangle = C'_{123} 
s_{12}^{a_3} s_{23}^{a_1} s_{13}^{a_2+1/2} \bigg[
  \frac{\theta_{12}}{s_{12}^{1/2}s_{23}^{-1/2}}  -
\frac{\theta_{23}}{s_{12}^{-1/2}s_{23}^{1/2}}\\
 - \frac{(h_{+,1} \theta_{23}
\theta_{12} \thetabar_{12} + (1-h_{+,1}-h_{+,2}) \theta_{12} \theta_{23}
\thetabar_{23})}{2s_{12}^{1/2}s_{23}^{1/2}} \bigg];
\end{multline}
when $H_+=1$ we have
\begin{multline}
\label{3pto2}
\langle 0 | \phi_1 \phi_2 \phi_3 | 0 \rangle = C''_{123} 
s_{12}^{a_3} s_{23}^{a_1} s_{13}^{a_2+1/2} \bigg[
  \frac{\thetabar_{12}}{s_{12}^{1/2}s_{23}^{-1/2}}  -
\frac{\thetabar_{23}}{s_{12}^{-1/2}s_{23}^{1/2}}\\
 - \frac{(h_{+,1} \thetabar_{23}
\theta_{12} \thetabar_{12} + (-1-h_{+,1}-h_{+,2}) \thetabar_{12} \theta_{23}
\thetabar_{23})}{2s_{12}^{1/2}s_{23}^{1/2}} \bigg].
\end{multline}
These are identical to the expressions given in \cite{Bl} and corrected from
\cite{MSS}.  Again, they are not invariant under permutations of the
field labels.  However, we find that (with the proviso discussed
above that anti-holomorphic coordinates should be included) the form
of these expressions for \{123\} is the same as that for \{321\}, {\em etc.}
 Indeed, if we sum the resulting three variants in each case, we obtain the
 solutions found by Howe and West \cite{West}.

To proceed with the calculation of fusion rules, we note that the
expression \eqref{schem1} is in terms of highest weight states rather
than fields acting on the vacuum.  We have the definition \eqref{stfi}
to give us the highest weight in state.  The global superconformal
transformations may be found by exponentiating the generators to be
\begin{align}
x'&=\frac{ax+b}{cx+d}+ \frac{e^q\theta((1-\hf\epsilon_1
\bar{\epsilon}_2)\bar{\epsilon}_1 x + (1+\hf\epsilon_2
\bar{\epsilon}_1)\bar{\epsilon}_2)}{(cx+d)^2} + 
\frac{e^{-q}\thetabar((1+\hf\epsilon_2 
\bar{\epsilon}_1)\epsilon_1 x + (1-\hf\epsilon_1
\bar{\epsilon}_2)\epsilon_2)}{(cx+d)^2}\notag\\
 &\qquad+
 \frac{\theta\thetabar((2d\epsilon_1\bar{\epsilon}_1-c(\epsilon_1\bar{\epsilon}_2+\epsilon_2\bar{\epsilon}_1))x+d(\epsilon_1\bar{\epsilon}_2+\epsilon_2\bar{\epsilon}_1)-2c\epsilon_2\bar{\epsilon}_2)}{(cx+d)^3},\notag\\
\theta'&=\frac{\epsilon_1 x+\epsilon_2}{cx+d} +
\frac{e^q\theta(1+\hf(\epsilon_2 \bar{\epsilon}_1-\epsilon_1\bar{\epsilon}_2)
  - \frac{1}{4}\epsilon_1\bar{\epsilon}_1\epsilon_2\bar{\epsilon}_2)}{cx+d} +
\frac{\theta\thetabar(d\epsilon_1-c\epsilon_2)}{(cx+d)^2},\notag\\
\thetabar'&=\frac{\bar{\epsilon}_1 x+\bar{\epsilon}_2}{cx+d} +
\frac{e^{-q}\thetabar(1+\hf(\epsilon_2
  \bar{\epsilon}_1-\epsilon_1\bar{\epsilon}_2)
  - \frac{1}{4}\epsilon_1\bar{\epsilon}_1\epsilon_2\bar{\epsilon}_2)}{cx+d} -
\frac{\theta\thetabar(d\bar{\epsilon}_1-c\bar{\epsilon}_2)}{(cx+d)^2},
\end{align}
here corrected from \cite{Kir} ($a$, $b$, $c$ and $d$ are the $SL(2)$
parameters $|ad-bc|=1$, $\epsilon$ and $\bar{\epsilon}$ are anticommuting 
parameters associated with the supersymmetry transformations and $q$ with the
transformation arising from $U_0$).

For a suitable definition of out state
$\langle \Lambda|$, we wish to take $x \rightarrow \infty$ via the
transformation
\begin{equation}
x'=\frac{1}{x}.
\end{equation}
For the global transformation to be of this form, we require that
$\epsilon_1=\epsilon_2=\bar{\epsilon}_1=\bar{\epsilon}_2=0$.  Consequently,
the transformations for $\theta$ and $\thetabar$ are given by
\begin{align}
\theta'=\frac{e^q\theta}{x},\notag\\
\thetabar'=\frac{e^{-q}\thetabar}{x}.
\end{align}
Then the point $(0,0,0)$ is mapped to $(\infty,0,0)$ as its natural
inverse, which is then the limit for the out state 
\begin{equation}
\langle \Lambda|= \lim_{\stackrel{\scriptstyle x\rightarrow
0}{\theta=\thetabar=0}} x^{h_-}\langle0|\phi\left(\frac{1}{x},
\frac{\theta}{x},\frac{\thetabar}{x}\right). 
\end{equation}
The factor $x^{h_-}$ arises from the transformation law for superprimary
fields \cite{Kir}
\begin{equation}
\tilde{\phi}(x,\theta,\thetabar) =
\phi(x',\theta',\thetabar')[(\partial_{\theta}+\hf
\thetabar\partial_x)\theta']^{(-h_--h_+)/2} [(\partial_{\thetabar}+\hf
\theta\partial_x)\thetabar']^{(-h_-+h_+)/2}
\end{equation}  
the factor in which reduces to $x^{h_-}$ for the transformation given, 
evaluated at $\theta=\thetabar=0$ and with the choice $q=0$.
 
Alternatively, we may consider the expansion of a primary field as given by
\begin{equation}
\phi(x,\theta,\thetabar)=\varphi(x)+\theta\psi(x)+
\thetabar\bar{\psi}(x)+\theta\thetabar g(x). 
\end{equation}
We see that with our previous definition of in state \eqref{stfi} we have 
\begin{equation}
\varphi(0)|0 \rangle = |\Lambda \rangle,
\end{equation}
so then
\begin{equation}
\langle \Lambda |=\lim_{x \rightarrow \infty}\langle 0 |\varphi(x)
x^{-h_-}
\end{equation}
with again $\theta=\thetabar=0$.

With the definition of out state established, we arrive at the form of
3-point function which we will use for our calculations of fusion
rules.  For the even case \eqref{3pte} for which
$h_{+,1}+h_{+,2}+h_{+,3}=0$, the result is 
\begin{equation}
\label{3ptel}
\langle \Lambda_1 | \phi (z_2,x_2,\theta_2,\thetabar_2) | \Lambda_3
\rangle = C_{123} z_2^{h_1-h_2-h_3} x_2^{(h_{-,2}+h_{-,3}-h_{-,1})/2}
\bigg[1-\frac{(h_{+,3}-2\alpha)}{2x_2}\theta_2\thetabar_2\bigg]. 
\end{equation}
For the odd case \eqref{3pto1} for which $h_{+,1}+h_{+,2}+h_{+,3}=-1$ we find
\begin{equation}
\label{3pto1l}
\langle \Lambda_1 | \phi (z_2,x_2,\theta_2,\thetabar_2) | \Lambda_3
\rangle = \tilde{C}'_{123} z_2^{h_1-h_2-h_3}
x_2^{(h_{-,2}+h_{-,3}-h_{-,1}-1)/2}\theta_2
\end{equation}
and the other odd case \eqref{3pto2}, where
$h_{+,1}+h_{+,2}+h_{+,3}=1$, becomes
\begin{equation}
\label{3pto2l} 
\langle \Lambda_1 | \phi (z_2,x_2,\theta_2,\thetabar_2) | \Lambda_3
\rangle = \tilde{C}''_{123} z_2^{h_1-h_2-h_3}
x_2^{(h_{-,2}+h_{-,3}-h_{-,1}-1)/2}\thetabar_2.
\end{equation}
The $z$-dependence is determined, as usual, by taking the commutator
with $L_0$: as it will not influence our discussion of fusion rules,
we shall generally omit it in what follows.

We should mention at this point that in the limit discussed above,
where $x_1\rightarrow \infty$, $x_3=0$ and
$\theta_1=\thetabar_1=\theta_3=\thetabar_3=0$, the odd 3-point
functions as given by Howe and West \cite{West} reduce to the
expressions \eqref{3pto1l} and \eqref{3pto2l}.  However, the
expression obtained by this procedure from their even 3-point
function differs from \eqref{3ptel}.  We will show that \eqref{3ptel}
leads to sensible fusion rules, whereas use of the corresponding Howe
and West expression only gives these in part.  Beyond this,
we can give no formal justification of why one might start from a
non-permutation invariant expression for the 3-point function
(which is thus intrinsically non-local).  We might also note that each
of the possible ways of writing \eqref{3pte} leads to the same
expression \eqref{3ptel}, that is
\begin{multline}
\lim_{\stackrel{\scriptstyle x_i \rightarrow
\infty,x_k=0}{\theta_{i,k}=0,\thetabar_{i,k}=0}} x_i^{-h_{-,i}}\langle 0 | \phi_i(x_i,\theta_i,\thetabar_i)
\phi_j(x_j,\theta_j,\thetabar_j) \phi_k(x_k,\theta_k,\thetabar_k) |
0 \rangle = \\
 C_{ijk}x_j^{(h_{-,j}+h_{-,k}-h_{-,i})/2}
\bigg[1-\frac{(h_{+,k}-2\alpha)}{2x_j}\theta_j\thetabar_j\bigg], 
\end{multline}
where $i,j,k=1,2,3$, $i\neq j\neq k$.

\section{Singular Vectors for $k=-1/2$}

In the Ramond sector of $\hslck$ at level $k=-1/2$, there are four
primary fields $\phi_{m,m'}$, $m,m'=0,1$ in the
so-called class $IV$ and class $V$ representations, these being the
relevant ones since the evidence indicates that their characters form
a closed set under modular transformations \cite{GBJ}.  The fields 
$\phi_{0,0}$ and $\phi_{0,1}$ are self-conjugate, while 
$\phi^*_{1,0}=\phi_{1,1}$.  In
\cite{BT97}, the authors calculated general expressions for singular
vectors using a Malikov-Feigin Fuchs type construction; we will make use of
these expressions here.  For each of the fields $\phi_{m,m'}$ there are three
singular vectors at the first level, from which conditions obtained through
the calculation of \eqref{schem1} must be simultaneously satisfied.  We shall
list the appropriate form of singular vectors as given in \cite{BT97} and go
on to make use of them to calculate fusion rules in the next section. 

For $|\Lambda_{0,0}\rangle$, with quantum numbers $h_-=h_+=h=0$, the three singular vectors
are 
\begin{align}
\label{la00}
(i)\qquad& j_0^-|\Lambda_{0,0}\rangle,\notag\\
(ii) \qquad& j_0^{'-}|\Lambda_{0,0}\rangle,\notag\\
(iii) \qquad& 
(J_{-1}^+)^{3/2}(j_0^-j_0^{'-}-j_0^{'-}j_0^-)
(J_{-1}^+)^{1/2}|\Lambda_{0,0}\rangle.  
\end{align}
For $|\Lambda_{1,0}\rangle$, $h_-=h_+=-\hf$, $h=0$ the singular vectors are
\begin{align}
\label{la10}
(i)\qquad& j_0^{'-}|\Lambda_{1,0}\rangle,\notag\\
(ii) \qquad& J_{-1}^+|\Lambda_{1,0}\rangle,\notag\\
(iii) \qquad& 
(J_0^-)^{1/2}(j_0^-j_0^{'-}-2j_0^{'-}j_0^-)J_{-1}^+j_0^-J_{-1}^+(J_0^-)^{-1/2}(-j_0^{'-}j_0^-)|\Lambda_{1,0}\rangle.
\end{align}
The state $|\Lambda_{1,1}\rangle$ has $h_-=-\hf$, $h_+=\hf$, $h=0$ and singular vectors
\begin{align}
\label{la11}
(i)\qquad& j_0^-|\Lambda_{1,1}\rangle,\notag\\
(ii) \qquad& J_{-1}^+|\Lambda_{1,1}\rangle,\notag\\
(iii) \qquad&
(J_0^-)^{1/2}(3j_0^-j_0^{'-} -J_0^-)
J_{-1}^+
j_0^{'-}J_{-1}^+(J_0^-)^{-3/2}(-j_0^-j_0^{'-}+J_0^-)|\Lambda_{1,1}\rangle 
\end{align}
and for $|\Lambda_{0,1}\rangle$ with $h_-=1$, $h_+=0$, $h=\hf$ the singular vectors are
\begin{align}
\label{la01}
(i)\qquad& (\thf j_{-1}^+ + j_0^{'-}J_{-1}^+)|\Lambda_{0,1}\rangle,\notag\\
(ii) \qquad& 
(j_0^-j_0^{'-}-j_0^{'-}j_0^-)|\Lambda_{0,1}\rangle,\notag\\
(iii) \qquad& 
(-\thf j_{-1}^{'+} + j_0^-J_{-1}^+)|\Lambda_{0,1}\rangle.
\end{align}

These expressions for singular vectors may be used in \eqref{schem} to give
expressions of the form \eqref{schem1}, utilising the equations  \eqref{fcr}.
The singular vectors generally involve fractional powers of generators, which
may be rearranged using 
\begin{equation}
\label{rear}
AB^a=\sum_{i=0}^{\infty}\binom{a}{i}B^{a-i}[\cdots[[A,\overbrace{B],B],\cdots,B]}^{i}
\end{equation}
to give expressions with integer powers.  For the purposes of calculation, it
is more convenient to keep the expressions as they stand and modify
\eqref{fcr} accordingly, using 
\begin{equation}
\label{fdcr}
\phi_j(x,\theta,\thetabar)(X_0)^a=\sum_{i=0}^{\infty}\binom{a}{i}(X_0)^{a-i}(-D^X_j)^i\phi_j(x,\theta,\thetabar),
\end{equation}
with an overall minus sign as required for the case of fermionic generators
and a fermionic field $\phi_j$.  This results in the differential operators
in \eqref{schem1} also involving fractional powers, in our case, of the
differential operators corresponding to the generators $J_0^-$ and $J_0^+$.
To deal with this situation, we will make use of the following expressions in
our calculations:
\begin{align}
(D^-)^{a}\theta^{\gamma}\thetabar^{\bar{\gamma}}x^b &=
(\partial_x)^{a}\theta^{\gamma}\thetabar^{\bar{\gamma}}x^b\notag\\ 
&= \frac{\Gamma(b+1)}{\Gamma(b-a+1)}\theta^{\gamma}
\thetabar^{\bar{\gamma}}x^{b-a};
\end{align}
\begin{align}
(D^+)^{a}\theta^{\gamma}\thetabar^{\bar{\gamma}}x^b &=
(-x^2\partial_x - x\theta\partial_{\theta} -x\thetabar\partial_{\thetabar} 
+xh_- +\hf \theta\thetabar h_+)^{a}\theta^{\gamma}\thetabar^{\bar{\gamma}}x^b \notag\\
&= \frac{\Gamma(h_- -b- \gamma-\bar{\gamma}+1)}{\Gamma(h_- -b-a- \gamma-\bar{\gamma}+1)}\theta^{\gamma}\thetabar^{\bar{\gamma}}x^{b+a} 
+\frac{ah_+\Gamma(h_- -b)}{2\Gamma(h_- -b-a+1)}\theta^{\gamma+1}\thetabar^{\bar{\gamma}+1}x^{b+a-1}.
\end{align}
These expressions may be verified as holding for integer values of $a$, with
the validity for fractional values of $a$ following by analytic continuation.

\section{Calculation of Fusions}

The information presented in the above sections allows us now to calculate
fusion rules.  As we wish to calculate expressions of the form
\eqref{schem1}, we note that since these are equal to zero, the procedure for
deriving \eqref{schem1} from \eqref{schem} essentially amounts to replacing the
generators by their corresponding differential operators, with any quantum
numbers involved in those expressions being the ones associated to the field
$\phi_2$, through which we are commuting.  We may ignore
factors of $z$ arising from those generators with mode numbers not equal to
zero and the possibility of having to use anticommutators between fields and
fermionic generators, since this only gives rise to an overall minus sign.
Consider  the singular vector $(iii)$ of \eqref{la00}.  Using
\eqref{fdcr} we have
\begin{multline} 
\langle \Lambda^*_1|\phi_2(z,x,\theta,\thetabar)
(J_{-1}^+)^{3/2}(j_0^-j_0^{'-}-j_0^{'-}j_0^-)
(J_{-1}^+)^{1/2}|\Lambda_3\rangle =\\
\langle \Lambda^*_1|(J_{-1}^+-z^{-1}D^+_2)^{3/2}(j_0^-
-d^-_2)(j_0^{'-}-d^{'-}_2)-\\
(j_0^{'-}-d^{'-}_2)(j_0^- -d^-_2)
 (J_{-1}^+-z^{-1}D^+_2)^{1/2}
 \phi_2(z,x,\theta,\thetabar)|\Lambda_3\rangle =0.
\end{multline}
This expression may be rearranged using \eqref{rear}, with the
generators arising from this procedure all such that they annihilate the
out state, as can be seen from the mode numbers involved.  The only
remaining part is the piece made up of the derivative terms, with an 
overall factor involving powers of $(-1)$ and $z$, which can
be eliminated.  The derivative terms may then be ``unrearranged'' to
give the expression with fractional powers and we have
\begin{multline}
\langle \Lambda^*_1|\phi_2(x,\theta,\thetabar)
(J_{-1}^+)^{3/2}(j_0^-j_0^{'-}-j_0^{'-}j_0^-)
(J_{-1}^+)^{1/2}|\Lambda_3\rangle \rightarrow\\ 
(D_2^+)^{3/2}(d_2^-d_2^{'-}-d_2^{'-}d_2^-)
(D_2^+)^{1/2}\langle\Lambda^*_1|\phi_2(x,\theta,\thetabar)
|\Lambda_3\rangle=0.    
\end{multline}
The only instance where some care needs to be taken is for the case of
$|\Lambda_{0,1}\rangle$, where there are singular vectors made up of a term involving one
generator and a term involving two generators, which will lead to a minus
sign difference on commuting with $\phi_2$.  It remains to apply each of the
three singular vectors for each field to the three 3-point functions
\eqref{3ptel}, \eqref{3pto1l} and \eqref{3pto2l}.

$\underline{\phi_{0,0}: h_-=h_+=0}$

We begin by examining $\phi_{0,0}$, the identity field in this context,
where we hope the behaviour to be fairly transparent.  For the even
3-point function \eqref{3ptel}, where now $H_+=h_{+,1}+h_{+,2}=0$ we calculate
\begin{align}
\label{sing} 
(i) \qquad& d_2^- \langle
\Lambda^*_1|\phi_2(x,\theta,\thetabar)|\Lambda_{0,0}\rangle =0\notag\\
(ii) \qquad& d_2^{'-}\langle
\Lambda^*_1|\phi_2(x,\theta,\thetabar)|\Lambda_{0,0}\rangle =0\notag\\
(iii) \qquad& (D_2^+)^{3/2}(d_2^-d_2^{'-}-d_2^{'-}d_2^-)(D_2^+)^{1/2} 
\langle \Lambda^*_1|\phi_2(x,\theta,\thetabar)|\Lambda_{0,0}\rangle =0,
\end{align}
with
\begin{equation}
\langle \Lambda^*_1|\phi_2(x,\theta,\thetabar)|\Lambda_{0,0}\rangle =
C[x^{a_1}+\hf(h_{+,1}+h_{+,2}+2\alpha)\theta \thetabar
x^{a_1-1}].
\end{equation}
From the first two equations, we find that
$a_1=\hf(h_{-,2}-h_{-,1})=0$ and $\alpha=0$.  Then from the third
equation we have two conditions: 
\begin{align}
&h_{-,2}(h_{-,2}-1)-3h_{+,2}^2=0,\notag\\
&h_{+,2}(h_{-,2}+\hf)=0.
\end{align}
When $h_{+,2}=0$ we have $h_{-,2}=0$ or $h_{-,2}=1$.  When
$h_{-,2}=-\hf$ we have  $h_{+,2}=-\hf$ or $h_{+,2}=\hf$.  This
unambiguously identifies the following possibilities: 
\begin{align}
\textrm{when }\phi_3=\phi_{0,0},\, \textrm{then }\quad& \phi_2=\phi_{0,0}
\textrm{ and } \phi^*_1=\phi_{0,0}\notag\\
\textrm{or }\quad& \phi_2=\phi_{1,0} \textrm{ and }
\phi^*_1=\phi_{1,1}\notag\\ 
\textrm{or }\quad& \phi_2=\phi_{1,1} \textrm{ and }
\phi^*_1=\phi_{1,0}\notag\\ 
\textrm{or }\quad& \phi_2=\phi_{0,1} \textrm{ and } \phi^*_1=\phi_{0,1}.
\end{align}

This is the sort of behaviour one would wish for, given that $\phi_{0,0}$
is the identity field.  However, when this calculation is performed
with the even 3-point function of Howe and West (with limits taken
as described), we only find the first and last of these results
arising, with no coupling between the identity and $\phi_{1,0}$ or
$\phi_{1,1}$. 

Repeating the exercise with the odd 3-point functions
\eqref{3pto1l} and \eqref{3pto2l} requires that these are identically
zero.  For example, considering case $(i)$ of \eqref{sing} using
\eqref{3pto1l} gives
\begin{equation}
(\partial_{\theta_2}-\hf\thetabar_2 \partial_{x_2})
\t{C}'_{123} x_2^{a_1-1/2}\theta_2 =
\t{C}'_{123}\left(x_2^{a_1-1/2}+\hf\left(a_1-\hf\right)\theta\thetabar
x_2^{a_1-3/2}\right) =0 
\end{equation}
which implies that $\t{C}'_{123}=0$.  Hence the even case exhausts all
possibilities.

$\underline{\phi_{0,1}: h_-=1,h_+=0}$

The next case we examine is that of $\phi_{0,1}$.  The even 3-point
function (with $h_{+,1}+h_{+,2}=0$) yields:
\begin{align}
\textrm{when } \phi_3=\phi_{0,1},\, \quad& \phi_2=\phi_{0,0}
\textrm{ and } \phi^*_1=\phi_{0,1}\notag\\
\textrm{or } \quad& \phi_2=\phi_{0,1} \textrm{ and } \phi^*_1=\phi_{0,0}.
\end{align}
The odd 3-point function \eqref{3pto1l}, for which
$h_{+,1}+h_{+,2}=-1$ gives:
\begin{equation}
\textrm{when } \phi_3=\phi_{0,1},\quad \phi_2=\phi_{1,0} 
\textrm{ and } \phi^*_1=\phi_{1,0} 
\end{equation}
while the other odd 3-point function ($h_{+,1}+h_{+,2}=1$) reveals:
\begin{equation}
\textrm{when } \phi_3=\phi_{0,1},\quad \phi_2=\phi_{1,1} 
\textrm{ and } \phi^*_1=\phi_{1,1}. 
\end{equation}

$\underline{\phi_{1,0}: h_-=-\hf,h_+=-\hf}$

Turning now to $\phi_{1,0}$ we notice that for this particular case of
quantum numbers, the singular vector in case $(iii)$ of \eqref{la10}
gives no additional information over case $(i)$.  Once the fact that
$j^{'-}_0|\Lambda_{1,0}\rangle=0$ has been imposed, case $(iii)$ vanishes
after the first step and this singular vector need not be considered.

For the even 3-point function, where $h_{+,1}+h_{+,2}-\hf=0$, we find:
\begin{equation}
h_{-,2}=-h_{+,2}=a_1=\hf \left(h_{-,2}+\left(-\hf\right)-h_{-,1}\right).
\end{equation}
While the quantum numbers of $\phi^*_1$ and $\phi_2$ are not given
explicitly, we can allow $\phi_2$ to take the quantum numbers of all
the $\phi_{m,m'}$ in turn and see what results this gives for
$\phi^*_1$.  In fact, since $h_{-,2}=-h_{+,3}$ we are immediately
restricted to taking $\phi_2=\phi_{0,0}$ or $\phi_2=\phi_{1,1}$.  Then
\begin{align}
\textrm{when } \phi_3=\phi_{1,0}, \textrm{ and } & \phi_2=\phi_{0,0}
\textrm{ then } \phi^*_1=\phi_{1,1}\notag\\ 
\textrm{and when } & \phi_2=\phi_{1,1} \textrm{ then }
\phi^*_1=\phi_{0,0} 
\end{align}
which is in agreement with results from the $\phi_{0,0}$ calculation 
(although with different values of the parameter $\alpha$).  

In the case of the odd 3-point function \eqref{3pto2l} we find
that this is identically zero.  However, the 3-point function
\eqref{3pto1l} (for which $h_{+,1}+h_{+,2}-\hf=-1$) gives
\begin{equation}
h_{-,2}=a_1+\hf = \hf -h_{-,1}.
\end{equation} 
Again, letting $\phi_2$ take the quantum numbers of $\phi_{m,m'}$ yields
\begin{align}
\textrm{when } \phi_3=\phi_{1,0},\textrm{ and }&\phi_2=\phi_{0,0}
\textrm{ then } h^*_{-,1}=\hf \textrm{ and }h^*_{+,1}=-\hf\notag\\
\textrm{ when }&\phi_2=\phi_{1,0} \textrm{ then } \phi^*_1=\phi_{0,1}\notag\\
\textrm{ when }&\phi_2=\phi_{1,1} \textrm{ then }  h^*_{-,1}=1
\textrm{ and } h^*_{+,1}=-1\notag\\ 
\textrm{ when } &\phi_2=\phi_{0,1} \textrm{ then } \phi^*_1=\phi_{1,0}.
\end{align}
There are two cases here for which the quantum numbers do not
correspond to any of the fields available.  However, these are
precisely the cases already considered in the even 3-point function
and so may be discarded.  The last result is as already obtained in
the consideration of $\phi_{0,1}$.

$\underline{\phi_{1,1}: h_-=-\hf,h_+=\hf}$

The situation for $\phi_{1,1}$ is very similar to that for
$\phi_{1,0}$.  The singular vector $(iii)$ of \eqref{la11} is of the
form $(\dots)(-j_0^-j_0^{'-}+J_0^-)|\Lambda_{1,1}\rangle$ which may be
rearranged as $(\dots)(j_0^{'-}j_0^-)|\Lambda_{1,1}\rangle$.  This will
again give no additional information over the result of using the
singular vector $(i)$ in \eqref{la11}, which is
$j_0^-|\Lambda_{1,1}\rangle$.  

The even 3-point function, with $h_{+,1}+h_{+,2}+\hf=0$ shows that
\begin{equation}
h_{-,2}=h_{+,2}=a_1=\hf\left(h_{-,2}+\left(-\hf\right)-h_{-,1}\right).
\end{equation}
We see that the only options for $\phi_2$ are $\phi_{0,0}$ and
$\phi_{1,0}$.  Hence
\begin{align}
\textrm{when } \phi_3=\phi_{1,1},\textrm{ and } & \phi_2=\phi_{0,0}
\textrm{ then } \phi^*_1=\phi_{1,0}\notag\\
\textrm{and when }  & \phi_2=\phi_{1,0} \textrm{ then } \phi^*_1=\phi_{0,0},
\end{align}
again with different values of $\alpha$.  As for the odd 3-point
functions, it is now \eqref{3pto1l} which is identically zero and
\eqref{3pto2l} (where $h_{+,1}+h_{+,2}=\hf=1$) that gives
\begin{equation}
h_{-,2}=a_1+\hf=\hf-h_{-,1}.
\end{equation} 
Considering the remaining options for $\phi_2$, we find
\begin{align}
\textrm{when } \phi_3=\phi_{1,1}, \textrm{ and }& \phi_2=\phi_{1,1}
\textrm{ then } \phi^*_1= \phi_{0,1}\notag\\
\textrm{and when }&\phi_2=\phi_{0,1} \textrm{ then } \phi^*_1=\phi_{1,1}.  
\end{align} 
The first of these results has already been seen in considering
$\phi_{1,0}$ while the second echoes the result of the $\phi_{0,1}$
calculation. 

To summarise the above results, replacing the fields $\phi^*_1$ by
their relevant conjugates, we have found that the following
fusion rules hold for the Ramond fields of $\hslck$ with $k=-\hf$:
\begin{xalignat}{2}
\label{fusr}
&\phi_{0,0} \times \phi_{0,0}=\phi_{0,0},&\phi_{1,0}& \times
\phi_{1,1}=\phi_{0,0},\notag\\ 
&\phi_{0,0} \times \phi_{1,0}=\phi_{1,0},&\phi_{1,0}& \times
\phi_{0,1}=\phi_{1,1},\notag\\ 
&\phi_{0,0} \times \phi_{1,1}=\phi_{1,1},&\phi_{1,1}& \times
\phi_{1,1}=\phi_{0,1},\notag\\ 
&\phi_{0,0} \times \phi_{0,1}=\phi_{0,1},&\phi_{1,1}& \times
\phi_{0,1}=\phi_{1,0},\notag\\ 
&\phi_{1,0} \times \phi_{1,0}=\phi_{0,1},&\phi_{0,1}& \times
\phi_{0,1}=\phi_{0,0}. 
\end{xalignat}
These fusion rules form an associative algebra, as they should.  One
immediately obvious statement about these results is that on
interchanging $\phi_{1,0}$ and $\phi_{1,1}$ the form of the fusion
rules is unchanged.   This precisely reflects what was discovered in
the investigation of modular invariants in \cite{GBJ}.
There we found the permutation invariants 
\begin{equation}
\label{perm}
Z=\sum \chi_{m,m'}\bar{\chi}_{\Pi(m,m'},
\end{equation}
involving
$\Pi (m,m')=(m,(m-m') \mod{u})$.  This permutation leaves $\phi_{0,0}$
and $\phi_{0,1}$ unchanged, but interchanges $\phi_{1,0}$ and
$\phi_{1,1}$ and so the modular invariant \eqref{perm} would seem to
be a consequence of the fusion rule automorphism (though we have not
explicitly established the form of fusion rules for the remaining sectors
involved in the modular invariant).

As for the Neveu-Schwarz sector, in superconformal field theory it has been 
found \cite{Gab} that the fusion of two Ramond fields yields a Neveu-Schwarz 
field, with the form of fusion rules preserved due to the sector isomorphism.  
By analogy, we expect to find fusion rules of the form above, but now with 
$\phi_i^{NS}\times \phi_j^R=\sum {N_{ij}}^k\phi_k^{NS}$: we should replace 
$\phi_1$ and $\phi_3$ in \eqref{fusr} by their Neveu-Schwarz counterparts.  
The confirmation of this behaviour remains a task for the future.

\section{Conclusion}

The study of conformal field theories based on affine algebras at
fractional level is one that has been tackled somewhat sporadically
over the last decade.  As yet, no absolute consensus has been reached
even for $\hat{sl}(2)$ as to whether these can actually define {\em bona
fide} conformal field theories in their own right.  However, the   
evidence does seem to suggest that this is possible; in any case, 
other models may be obtained through hamiltonian reduction or the
coset construction.  The work of \cite{ER} is a first indication that
fusion rules are well-defined for fractional level superalgebras, a
conclusion which is also borne out by this work.  The authors of
\cite{ER} were able to determine consistent fusion rules for all
levels at which admissible representations of $\widehat{osp}(1|2)$
exist.  Due to the far more complex nature of singular vectors of
$\hslck$, we have thus far only examined a particular case, that of
$k=-\hf$, but feel that a complete examination of levels $k=1/u-1$,
$u \in \N \setminus \{1\}$ may not be out of the question.  However, for
the situation at
hand, we have found that the Ramond fields present do close under
fusion.  Other methods, particularly that of the Coulomb gas formalism 
(used in \cite{MSS} for $N=2$ superconformal field theory) would seem to 
be more promising for a more thorough examination of this problem of fusion 
rules.  As yet, the techniques for such a study have not been developed for 
superalgebras other than $\widehat{osp}(1|2)$ \cite{Ras}.
It will be interesting to deepen the study of results in
this area and compare this with the work of \cite{GBJ} on modular
transformations of $\hslck$ in the attempt to fully realise a conformal
field theory based on fractional level $\hslc$.

\subsection*{Acknowledgements}

GBJ thanks Peter Bowcock for useful discussions, and acknowledges the award 
of an EPSRC research studentship.  The Grassmann package for Maple VI 
\cite{CT} was very helpful in some computer calculations.

\def\NPB{Nucl.\ Phys.\ B } 
\def\CMP{Commun.\ Math.\ Phys.\ }


\begin{thebibliography}{[99]}


\bibitem{MW} P. Mathieu and M.A. Walton, Prog.\ Theor.\ Phys.\ Supp.\
102 (1990) 229.
\bibitem{AY} H. Awata and Y. Yamada, Mod.\ Phys.\ Lett.\ A7 (1992) 1185.
\bibitem{BF} D. Bernard and G. Felder, \CMP 127 (1990) 145.
\bibitem{FGP} P. Furlan, A.Ch. Ganchev and V.B. Petkova, \NPB 491
(1997) 635, hep-th/9608018.
\bibitem{FM} B. Feigin and F. Malikov, {\em Modular Functor and
Representation Theory of $\hat{sl}(2)$ at a Rational Level},
q-alg/9511011.
\bibitem{DLM} C.Y. Dong, H.S. Li and G. Mason, \CMP 184 (1997) 65.
\bibitem{FGP1} P. Furlan, A.Ch. Ganchev and V.B. Petkova, \NPB 518 [PM] (1998) 
645, hep-th/9709103.
\bibitem{FGP2} P. Furlan, A.Ch. Ganchev and V.B. Petkova, \CMP 202 (1999) 701, 
math.QA/9807106.
\bibitem{GPW} A.Ch.\ Ganchev, V.B. Petkova and G.M.T. Watts, \NPB 571
(2000) 457, hep-th/9906139.
\bibitem{PRY} J.L. Petersen, J. Rasmussen and M. Yu, \NPB 481 (1996)
577, hep-th/9607129.
\bibitem{And} O. Andreev, Phys.\ Lett.\ B 363 (1995) 166, hep-th/9504082.
\bibitem{ER} I.P. Ennes and A.V. Ramallo, \NPB 502 (1997) 671,
hep-th/9704065.
\bibitem{Sem} A.M. Semikhatov, \NPB 478 (1996) 209.
\bibitem{MS} Z. Maassarani and D. Serban, \NPB 489 (1997) 603, hep-th/9605062.
\bibitem{FZ} V.A. Fateev and A.B. Zamolodchikov, Sov.\ J.\ Nucl.\
Phys.\ 43 (1986) 657.
\bibitem{BS} M. Bauer and N. Sochen, \CMP 152 (1993) 127, hep-th/9201079.
\bibitem{Kir} E.B. Kiritsis, Phys.\ Rev.\ D 36 (1987) 3048.
\bibitem{MSS} G. Mussardo, G. Sotkov and M. Stanishkov, Int.\ J.\
Mod.\ Phys.\ A 4 (1989) 1135.
\bibitem{West} P. West, {\em Introduction to Supersymmetry and
Supergravity, Extended Second Edition}, (World Scientific, Singapore
1990); P. Howe and P. West, Phys.\ Lett.\ B 227 (1989) 395.
\bibitem{Bl} R. Blumenhagen, \NPB 405 (1993) 744, hep-th/9208069.
\bibitem{BT97} P. Bowcock and A. Taormina, \CMP 185 (1997) 467, hep-th/9605220.
\bibitem{BHT98} P. Bowcock, M.R. Hayes and A. Taormina, \NPB 510 [PM] (1998) 739, hep-th/9705234.
\bibitem{Ras} J. Rasmussen, \NPB 510 (1998) 688, hep-th/9706091.
\bibitem{Watts} G.M.T. Watts, \NPB 407 (1993) 213, hep-th/9306034.
\bibitem{GBJ} G. Johnstone, \NPB 577 (2000) 646, hep-th/9909067.
\bibitem{Gab} M.R. Gaberdiel, Int.\ J.\ Mod.\ Phys.\ A12 (1997) 5183,
hep-th/9607036. 
\bibitem{CT} E.S. Cheb-Terrab, {\em Symbolic Computing with Grassmann 
Variables}, hep-th/9510226.

\end{thebibliography}
\end{document}